\documentclass[]{aastex701}
\usepackage{amsmath, amssymb, graphics}
\usepackage{graphicx}
\usepackage{caption}
\usepackage{enumerate}   
\def\cc {\ifmmode{{\rm cm}^{-2}}\else{${\rm
cm}^{-2}$}\fi}
\def\ccc {\ifmmode{{\rm cm}^{-3}}\else{${\rm
cm}^{-3}$}\fi}

\def\kms  {km\,s$^{-1}$}

\def\aco {\ifmmode{^{12}{\rm CO}(J=1-0)}\else{$^{12}{\rm
CO}(J=1-0)$}\fi}
\def\bco {\ifmmode{^{12}{\rm CO}(J=2-1)}\else{$^{12}{\rm
CO}(J=2-1)$}\fi}
\def\cco {\ifmmode{^{12}{\rm CO}(J=3-1)}\else{$^{12}{\rm
CO}(J=3-2)$}\fi}

\def\gcs {\ifmmode{{\rm CS}(J=7-6)}\else{${\rm
CS}(J=7-6)$}\fi}

\def\ahcn {{${\rm
HCN}(J=1-0)$}}

\def\dhcn {{${\rm
HCN}(J=4-3)$}}

\def\Msun{\ifmmode{{\rm M}_{\odot}}\else{{M$_{\odot}$}}\fi}
\def\deg {$^{\circ}$}

\def\arcsec {$^{\prime\prime}$}

\newcommand\sgra {Sgr~A$^*$}

\let\a=\alpha \let\b=\beta \let\g=\gamma \let\d=\delta \let\e=\epsilon
\let\h=\eta  \let\k=\kappa \let\l=\lambda 
\let\s=\sigma \let\t=\tau \let\u=\upsilon
\let\r=\rho
 
\let\w=\omega	   \let\G=\Gamma \let\D=\Delta  \let\L=\Lambda
\let\S=\Sigma \let\W=\Omega

\def\be{\begin{eqnarray}}
\def\ee{\end{eqnarray}}
\def\ba{\begin{array}}
\def\ea{\end{array}}

\def \erg {{\rm erg}}

\def \yr {\mathrm{yr}}
\def \cm {\mathrm{cm}}
\def \M  {\mathscr{M}}
\def \K {\mathrm{K}}
\def \sec {\mathrm{s}}

\usepackage{xcolor}

\begin{document}

\title{The Discovery of an Active Wind from the Milky Way's Central Black Hole}

\author[orcid=0000-0001-9300-354X,sname='Mark Gorski']{Mark D. Gorski}
\affiliation{Center for Interdisciplinary Exploration and Research in Astrophysics, Northwestern University}
\email[show]{mark.gorski@northwestern.edu}  

\author[orcid=0000-0001-8986-5403, sname='Elena Murchikova']{Lena Murchikova} 
\affiliation{Center for Interdisciplinary Exploration and Research in Astrophysics, Northwestern University}
\affiliation{Department of Physics and Astronomy, Northwestern University}
\affiliation{School of Natural Sciences, Institute for Advanced Study, Princeton}
\email[show]{lena@northwestern.edu \\ Contributions of authors is equivalent.}

\begin{abstract}

Every large galaxy has a black hole in its center. The interaction between the black hole and its host profoundly shapes galactic evolution and the Universe as a whole.
The key features of this interaction are black hole jets -- or more generally, winds -- which every black hole must have. 
Despite the proximity and importance of our Galaxy's central black hole, Sagittarius A* (\sgra), the active wind from it has eluded scientists for over half a century. 
Here we report the discovery of a large active wind from \sgra \ using unprecedentedly deep (T$_{\rm{b}} \sim30$ mK) and high angular resolution ($\lesssim$0.25\arcsec) observations with the Atacama Large Millimeter/Submillimeter Array (ALMA).
We detect a large conical clearing in the cold molecular gas surrounding \sgra\ that is at least 1~parsec long and has a $\sim$45~degree opening angle.
The morphology and energetics of this structure are consistent with active clearing of gas by a hot wind from \sgra. 

\end{abstract}




\section{Introduction}

Supermassive black holes (SMBHs) reside at the centers of all large galaxies  \citep{Kormendy2013}. 
They play a crucial role in galaxy evolution and shape the Universe as a whole \citep{Silk&Rees1998,2018IllustrisTNG}.
Descriptions of black hole feeding -- accretion -- and a black hole's reaction to being fed -- feedback -- involve a combination of physical processes in radiation, plasma, atomic, molecular, and particle physics, and usually require computer simulations to study in quantitative detail \citep{Davis&Tchekho_AnnualReviews}.
Qualitatively, however, these processes can be explained relatively simply.
SMBHs are surrounded by deep gravitational potential wells filled with gas, stars, and dust. 
Gas is the primary fuel source for SMBH growth \citep{2002apa..book.....F}. 
It is abundant and easily loses angular momentum through friction and collisions, causing it to fall inward toward the black hole.
As the gas approaches the black hole, some of its potential energy is converted into thermal energy.
So SMBHs are expected to be surrounded by progressively hotter gas the closer one approaches to the black hole.
This heated material emits electromagnetic radiation -- called black hole emission -- which can reach up to 39\% of $\dot{M} c^2$ of the gas mass crossing the horizon, though typical values are much lower \citep{1973NovikovThorne,2014riafs}. 
SMBHs also drive winds, as only one gram of accreting gas can produce enough energy to unbind $\sim 100$ kg of gas at $0.5 \times 10^5$ gravitational radii from the black hole. 
These winds -- which can be called jets if they are collimated -- are expected to originate close to the black hole horizon, and therefore should be hot \citep{1984jet,1977BZ,Blandford2019}.
Black hole radiation and winds vary significantly depending on the black hole's intrinsic characteristics and local environmental conditions. 
However, as black hole feeding cannot cease, neither can the winds \citep{2002apa..book.....F}.

\begin{table*}[ht]
    \centering
    \caption{The ALMA observations used for analysis. Project code number, the date of the observation, angular resolution in arcseconds,  maximum recoverable scale in arcseconds (calculated from the 5th percentile of baselines), time on the target source \sgra,  time resolution used for modeling variability of the source for the calibrations with \texttt{UVMultiFit}, the included spectral window (SPWs), and channel width.}
    \label{tab:allobservations}
    \begin{tabular}{l c c c c c c c}
    \\
    \hline
        Project Code    & Date            & Ang. Res. & Max. Rec & Time on & Model Time & Spw(s) Range & Channel Width\\ 
                & yy-mm-dd         &  (arcsec) &Scale (arcsec) & \sgra \,(s) & Res (s) & GHz & MHz \\ \hline
        2016.1.00870.S  & 2017-05-09      & 0.172 & 4.10    & 4801  & 30    &  229.43-231.43    & 15.625     \\
                        & 2017-07-02      & ``    & ``      & 4729  & 30    & ``                & `` \\ 
                        & 2017-07-03      & ``    & ``      & 4379  & 30    & ``                & `` \\
                        & 2017-07-04      & ``    & ``      & 4387  & 30    & ``                & `` \\
                        & 2017-07-05      & ``    & ``      & 9118  & 30    & ``                & `` \\
                        & 2017-07-06      & ``    & ``      & 9830  & 30    & ``                & `` \\ \hline
        2017.1.00995.S  & 2018-03-13      & 0.428 & 6.68    & 5825  & 15    &  228.55-230.53    & 15.625     \\
                        &                 &       &         &       &       &  230.25-232.24    & 15.625     \\
                        & 2018-03-15      & ``    & ``      & 18526 & 15    & ``                & `` \\
                        & 2018-04-16      & 0.670 & 10.0    & 16904 & 15    &  228.55-230.53    & 15.625     \\ 
                        &                 &       &         &       &       &  230.25-232.24    & 15.625     \\
                        & 2018-04-18      & ``    & ``      & 5655  & 15    & ``                & `` \\
                        & 2018-04-20      & ``    & ``      & 7160  & 15    & ``                & `` \\ \hline               
        2019.1.01559.S  & 2021-07-20      & 0.113 & 2.39    & 5109  & 15    & 228.56-230.43    & 1.938       \\
                        &                 &       &         &       &       & 230.24-232.11    & 1.938       \\
                        & 2021-07-22      & ``    & ``      & 5137  & 15    & ``                & `` \\ \hline      
    \end{tabular}
\end{table*}

Our closest SMBH, Sagittarius A* (\sgra), residing in the center of the Milky Way, is our best opportunity to study the physics of galactic center black holes.
It is in a quiescent state, which represents the dominant phase in the life cycle of SMBHs that governs their long-term impact on galaxy evolution \citep[][and references therein]{Genzel2010, Markoff2010, Kormendy2013}
Surprisingly, \sgra\ seems to have no currently active jet or wind.
The wind from \sgra was first proposed in 1971 shortly after the black hole was first discovered \citep{Lynden-Bell1971}, and on large scales  ($100 - 10,000$~pc), there is abundant evidence of its past activity, usually perpendicular to the Galactic plane \citep{2010FermiBubbles,2020eRositaBubbles,2019XrayChimneys,Morris2010}.
On small scales of a few pc or less, however, evidence for a current or recent wind remains inconclusive and directionally conflicting. 
The proposed wind/jet orientations are approximately along the Galactic plane and on the plane of the sky \citep{Yusef-Zadeh2020,Peissker2020}, approximately perpendicular to the Galactic plane \citep{Li2013,2021CecilJet,Baganoff2003}, and along our line of sight at 30 degrees due to the black hole spin orientation \citep{EHT2022,Gravity_hotspot2018}.
It is fair to say that, to date, there is no universally accepted signature of a presently active wind from \sgra. 
This work rectifies this issue.

\section{Observations}
\subsection{Observations and Image Quality}
We combined data from multiple epochs of ALMA Band 6 observations of \sgra\ (project codes: 2016.1.00870.S, 2017.1.00995.S, and 2019.1.01559.S; PI: Murchikova).
Properties of each data set are collated in Table \ref{tab:allobservations}.
All observations were obtained at a frequency of about 230 GHz. 
The data reduction and imaging were performed in CASA v6.5.3 \citep{CASATeam2022PASP}.
The combined data set has a recoverable scale of $10.0$\arcsec, as measured from the 5th percentile of baselines from the observation with the most compact array configuration, and an angular resolution of $\sim0.2$\arcsec.
The absolute minimum baseline length in the combined data set is 12.4~m corresponding to a maximum recoverable angular scale of 26.4\arcsec.
Images were produced using Briggs weighting with a robust parameter of 0.5.
The channel widths are 15.625~MHz, corresponding to a velocity resolution of approximately 20~\kms\ at 230 GHz.

\begin{figure*}[!htb]
    \centering
    \includegraphics[width=\textwidth]{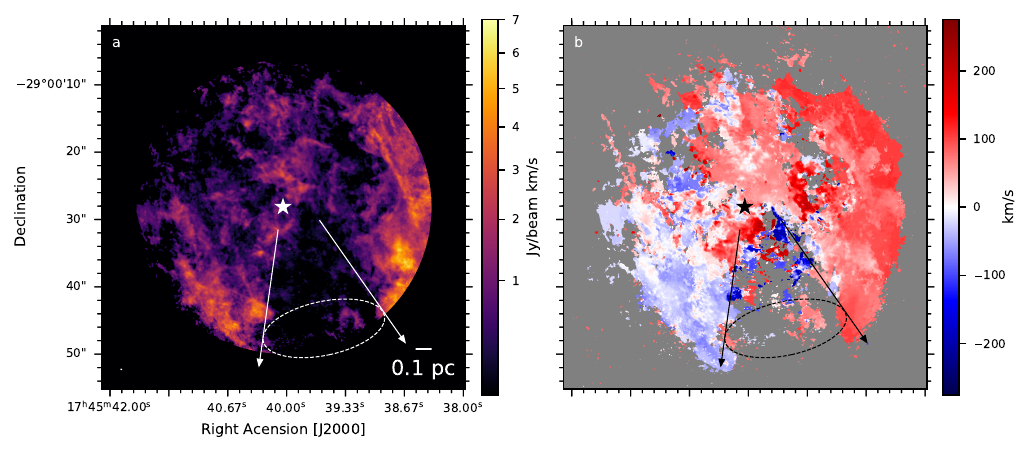}
    \caption{\bco \, line map in the inner 1~pc of the Milky Way. \textbf{a,} Integrated-flux map showing the intricate molecular gas structures. 
    The map is primary-beam-corrected. 
    Only pixels with a signal-to-noise ratio $>10\sigma$ were used within the velocity range of -195 \kms to 205 \kms. 
    Absorption towards the continuum sources, e.g., IRS~13, is not included.
    \textbf{b,} Velocity field of the \bco\ emission showing gas rotation around the black hole. 
    The location of \sgra \, is indicated with stars.
    The \sgra\ wind cone is marked with arrows.}
    \label{fig:COpb_vel}
\end{figure*}

The final data products were produced using CASA’s \texttt{tclean} task with the multiscale deconvolver.
We used scale sizes of 0, 7, 42, and 126 pixels and a Briggs weighting parameter of 0.5.
The pixel size is 0.02\arcsec, and the final image is 2,700$\times$2,700 pixels for a field of view of 54\arcsec\ (2~pc at a distance of 8~kpc \citealp{Do2019,Gravity2019}).
This encompasses  almost the entire primary beam  out to the first null at 232~GHz (54.2~\textrm{arcsec}).
Data cubes of the \bco\ line were produced with a spectral resolution of 25~\kms\ and have an RMS noise level of 80~$\mu$Jy.
The cube was smoothed to a common beam size of $0.256\, \textrm{arcsec} \times 0.235\, \textrm{arcsec}$, with a postilion angle of 60.3 degrees.
The resulting image is two orders of magnitude improved in sensitivity and a factor of 80 higher in resolution than previously published maps \citep{Marshall1994,Goicoechea2018_HVMC}.
\citet{Gorski2025} showed that absorption towards the CND in this line of sight is minimal.
The flux recovered is consistent with the known and expected properties of the CND.
The observations are also consistent with single-dish measurements of the Galactic Center.
Some artifacts remain within the inner 0.5\arcsec (0.002~pc radius) due to an incomplete subtraction of the black hole emission.
This region is masked in the analysis.
All velocities are reported in the LSRK frame using the radio convention.
Continuum subtraction was performed in the image domain by fitting a linear function to line-free channels.

\subsection{Data Calibration and Analysis}

We construct a time-variable model by fitting visibilities using the software \texttt{UVMultiFit} \citep{marti-vidal2014}.
We phase-only self-calibrate using a point source model of \sgra\ of constant 1 Jy at the phase center. \sgra\ was modeled as a delta function with a fixed position at the phase center.
This reduces phase errors such that the majority of the flux associated with \sgra\ is concentrated at \sgra's position.
We then derive a time-variable model of \sgra, using the phase-only corrected visibilities, with a time resolution of 15 or 30 s, using \texttt{UVMultiFit}. 
\sgra\ was again modeled as a delta function with a fixed position at the phase center, but with a variable amplitude per spectral channel. 
We found these time intervals were the shortest intervals for \texttt{UVMultiFit} to converge to a solution.
We applied phase and amplitude self-calibration using this time- and channel-dependent model to achieve further improved visibilities.
The model was subsequently subtracted from the calibrated visibilities to enable imaging of the surrounding Galactic Center.
Subtraction of \sgra\ is crucial to achieve high-dynamic-range images of the fainter structures around the black hole.  
Essentially, we generated a model of \sgra\ every $\sim$15~s by fitting a point source to the visibilities.
This model is used to phase and amplitude self-calibrate the data and then subtracted from the visibilities. 
After subtraction, we inspected the residuals to identify time intervals where the model may have failed to capture \sgra's variability or structure.
Time ranges showing variability inconsistent with a stable source were flagged and removed.
The large gaseous structures surrounding \sgra, e.g., the minispiral, are not expected to vary on hour-long timescales.
After this final round of flagging, the data were imaged using CASA's \texttt{tclean} task.

\begin{figure*}[!htb]
    \centering
    \includegraphics[width=\textwidth]
    {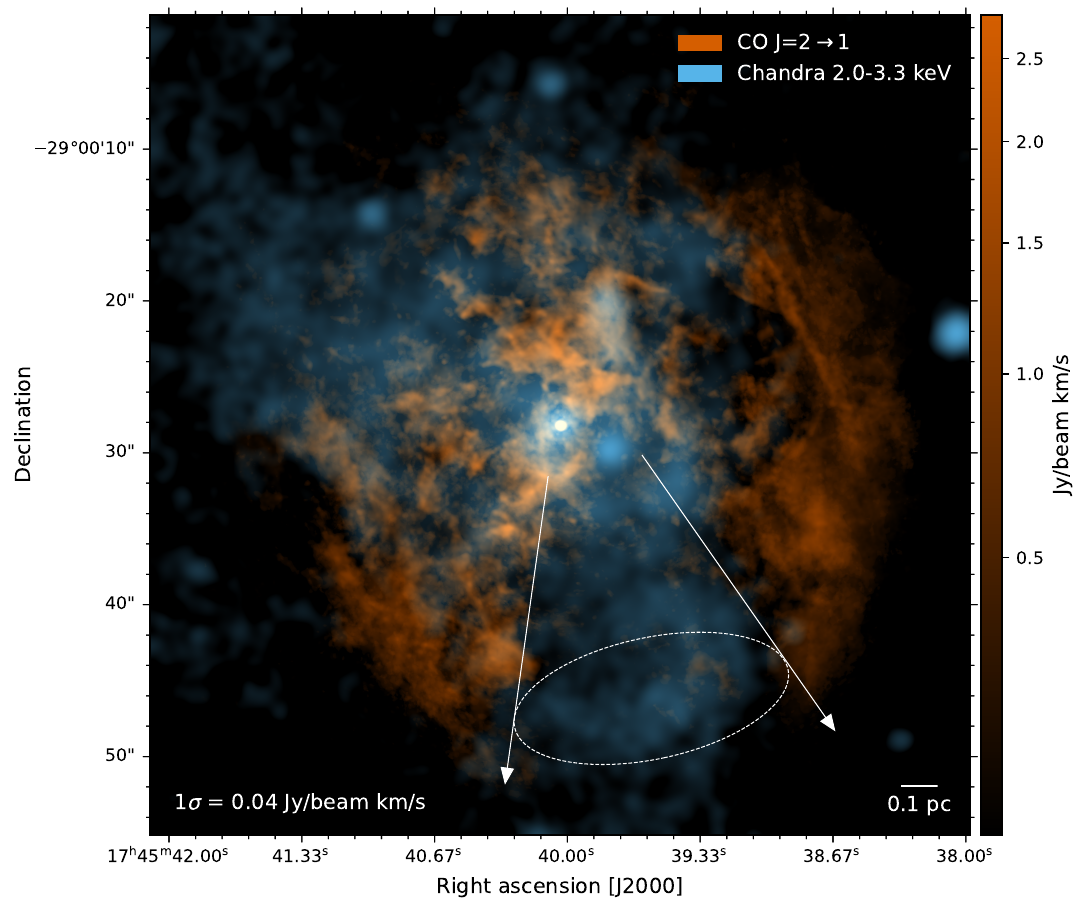}
    \caption{Molecular gas and X-ray emission around \sgra \, black hole. 
    ALMA observations of the \bco\ line at 230.538 GHz (orange) show $\sim$100 K molecular gas. 
    Chandra observations \citep{Muno2009,Morris2010} of 2.0-3.3 keV X-ray emission (blue) show hot $\sim 10^7$ K gas. 
    \sgra \, is a bright point in the middle of the image (most of its emission has been removed to improve imaging quality). 
    The image shows anti-correlation between CO and X-ray emission.
    The active black hole wind manifests as a large cavity nearly free from molecular gas emission and is marked with white arrows. 
    The ALMA \bco\ image has an angular resolution of $\sim$0.26\arcsec and is not primary-beam-corrected.
    $0.26$\arcsec is equivalent to 0.01~pc at the distance of the Galactic Center (D = 8~kpc )
    Chandra has a native resolution of $\sim0.5$\arcsec.}
    \label{fig:CO_Xray}
\end{figure*}

\section{Cold Molecular Gas Around SGR A*}\label{sec:cold_gas}

It is well established that the inner few pc of the Galactic Center contains a large amount of molecular gas ($\sim10^4$\Msun) \citep{Guesten1987,Montero-Castano2009,Oka2011,Etxaluze2011,Martin2012,Requena-Torres2012,Lau2013,2015Harada_CND,James2021,Hsieh2021,Tanaka2021,Christopher2005} with temperatures from several $10$ K to $\sim 500$~K \citep{James2021,Requena-Torres2012,Hsieh2021}.
The vast majority of this gas is contained in the Galactic Center molecular ring (Circumnuclear Disk or CND), which extends to $\sim3$-$5$~pc from the black hole. 
It was previously thought that the inner edge of the CND is about $0.5$~pc from \sgra, and the region inside this radius is filled mostly with atomic and hot ionized gas and devoid of cold molecular gas \citep{Genzel2010,2015Harada_CND, Jackson1993}. 
Below we describe our observations that reveal an abundance of intricate cold-gas structures within this region.


Carbon monoxide ($\rm{CO}$) is the most commonly used tracer of the molecular gas in the interstellar medium (ISM). 
It emits bright lines and is readily excited at the low densities and temperatures \citep{Bolatto2013b}.
The \bco\ transition is a rotational transition at the frequency of 230.538~GHz (the wavelength of 1.3 mm) 
which is efficiently excited and highly emissive at gas temperatures of $\sim 100$~K.

\begin{figure*}[!htb]
    \centering 
    \includegraphics[width=\textwidth]{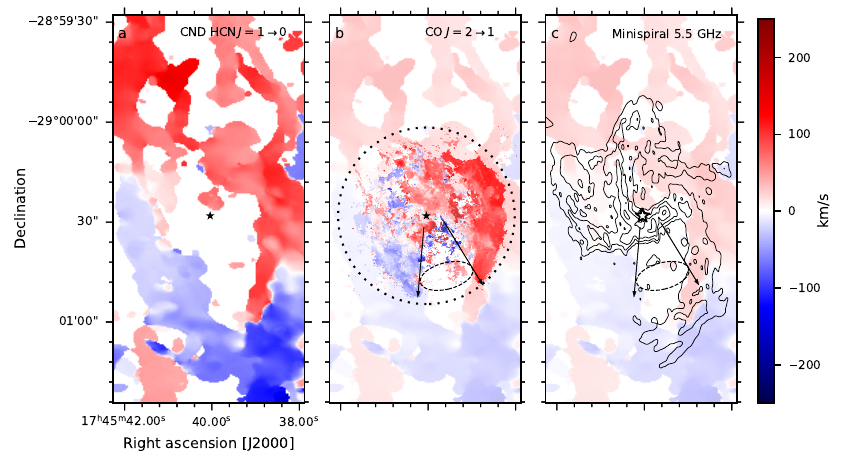}
    \caption{Cold molecular gas of the Galactic Center in the context of other structures.
    Velocity fields presented here are the intensity-weighted velocity using pixels with intensities $>10\sigma$ and the same velocity range as Figure \ref{fig:COpb_vel}.
    \textbf{a,} The CND shown in the velocity field of the \ahcn\ molecular transition \citep{Christopher2005,Hsieh2021} revealing the gas rotation. 
    \textbf{b,} Our \bco \, velocity map overlaid on the semi-transparent CND map of panel a. We see complete agreement, except our \bco\, map is more detailed. The black dotted circle shows the first null of the ALMA primary beam.
    The conical region cleared of molecular gas by \sgra's wind is marked with arrows to the south.
    \textbf{c,} contour map of the Galactic Center minispiral \citep{Zhao2016}, overlaid with the semi-transparent CND map of panel a
    and \sgra \, wind markers from panel b. The wind boundaries, if extended to the south, overlap with the Western Arc of the minispiral.
    The stars indicate the location of \sgra. 
    }
    \label{fig:CND_vels}
\end{figure*}

We present the highest-resolution and most sensitive map of cold gas within $\sim 1$~pc from \sgra \ to date: 
Figure \ref{fig:COpb_vel} is 100 times deeper than the previous molecular gas map \citep{Goicoechea2018_HVMC} and  80 times sharper than the previous \bco \, map of this region \citep{Marshall1994}.
The individual channel maps are shown in Appendix~\ref{app:known_structures}.

The achieved high imaging depth and resolution allowed us to see the following:
\begin{enumerate}[(i)]

\item The interior of the CND, which previously was believed to be hot, nearly free from 100~K gas, and having a sharp transition between hot and cold gas phases \citep{Genzel2010}, is instead filled with cold molecular gas that  is continuously connected to previously observed massive gas structures 
(Figure \ref{fig:COpb_vel}-\ref{fig:CO_Xray}).

\item The gas shows an ordered flow with a distinct redshifted and blueshifted velocity pattern, characteristic of rotation and/or infall toward \sgra\ (Figures \ref{fig:COpb_vel}b and \ref{fig:CND_vels}). Though the velocity range of the data cube does not cover velocities greater than the escape speed $\gtrsim\pm750$~\kms of the central pc.

\item The distribution and velocity structure of the brightest emission we detected in the 100~K gas are fully consistent with previous observations of the CND (Figure \ref{fig:CND_vels}ab) and other known structures (Appendix \ref{app:known_structures}). 
The match is particularly striking in Figure \ref{fig:CND_vels}ab.
The CO map we presented matches the map of the CND in the known tracer of the molecular gas in the central parsec HCN largely one-to-one, except with higher spatial resolution. 
Because a foreground or background structure with nearly identical geometry and velocity would be highly improbable, we conclude that the emission indeed originates from the immediate Galactic Center.

\item The $100$~K molecular gas emission anti-correlates with X-ray emission (Figures \ref{fig:CO_Xray}, and a detailed discussion in Appendix \ref{app:known_structures}), as expected, because the presence of hot $\sim 10^7$ K X-ray-emitting plasma implies the lack of cold gas in the same region \citep{Maloney1996}. 

\item The cold gas inside the commonly assumed inner edge of the CND is clumpy, with average density estimated to be just under the lowest density gas regions within the CND itself. However, the estimate is uncertain  (Appendix \ref{app:wind energy}).

\item Finally, and crucially for this work, we identify a large (in its footprint in the inner $\sim 1$ pc of the Galactic Center) conical clearing in the cold gas (Figure \ref{fig:CO_Xray} marked with arrows), with an opening angle of about 45 degrees and spanning the full extent of the image $\sim 1$~pc.
The edges of the cavity are consistent with models of the filaments in the Galactic Center proposed by \citet{Muzic2007}.
We argue below that this cavity is created by a hot active wind originating from \sgra.

\end{enumerate}

Part of our CO map includes emission from +80 \kms\ feature in the western edge of the image.
Although the cloud's exact trajectory is unknown, it is known to interact with the CND and OH streamer within the central parsec of the Galaxy \citep{Karlsson2015}.
It is considered to be part of the CND.
Depending on CND and +80 \kms \, cloud's relative geometry, the wind cone opening angle is $\sim 45$ degrees or possibly slightly larger.
The outer lobes of the CND lie outside our field of view \citep[e.g.][]{Yusef-Zadeh2001}.

\section{Evidence of the Wind}\label{sec:outflow}

Black hole winds originate near the event horizon, where gas temperatures are high.
As the wind travels outwards, it is expected to clear the cold molecular gas on its path.
The CND is permeated by low temperature molecular gas \citep[e.g.,][and references above]{Tanaka2021}.
In this context an outflow/wind from the black hole would be identified by a clearing of cold gas on the path of the high-temperature plasma.
We use the term ``wind" here, as it is not collimated enough to be called a jet \citep{1984jet}.
Figures \ref{fig:COpb_vel} and \ref{fig:CO_Xray}, show a cone devoid of cold gas with a 45-degree opening angle (Section \ref{sec:cold_gas} (vi)) extending from \sgra\ to the edge of our field of view ($\sim1$~pc). 
We argue that the observed CO cavity is likely 
created by a hot active wind originating from \sgra.

\subsection{Evidence of the Southern Lobe}

The orientation of the cavity relative to the CND suggests that the wind is directed on the sky toward the south-southwest (SSW). 
This interpretation is consistent with near‑infrared opacity measurements \citep{Muzic2007,Schodel2010,Lau2013} and radio measurements of Galactic Center filaments \citep{Zadeh2017a,Yusef-Zadeh2023}. 
Low‑opacity regions are found along the wind axis in both directions, and several filaments have been previously attributed to outflows from \sgra.

\begin{figure}
    \centering
    \includegraphics[width=\columnwidth]{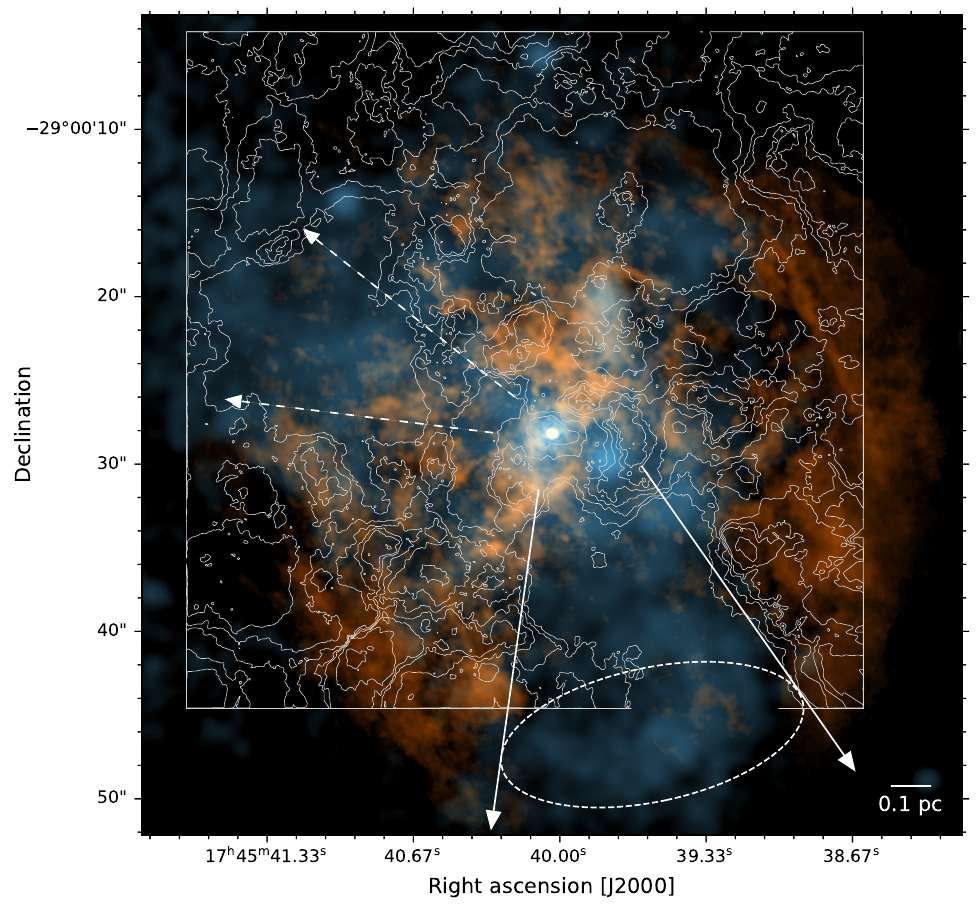}
    \caption{Same as Figure \ref{fig:CO_Xray}, but with the near-infrared extinction map contours showing the local extinction in the central pc from \citet{Schodel2010}.
    Contours start at 2.6 magnitudes, increasing in 0.1 magnitude steps until the maximum extinction in the map of 3.2 magnitudes. 
    A low-opacity cavity to the south-southwest on the sky (solid arrows) is coincident with the diffuse X-ray emission and the cavity traced by \bco. 
    A tentative counter wind cone is identified in the NNE direction (dashed arrows).}
    \label{fig:Extinct}
\end{figure}

We find agreement between the direction of \sgra's SSW wind and the Western Arc of the Galactic Center minispiral \citep{Lo1983} (Figure \ref{fig:CND_vels}c).
Although the end of the wind cone lies beyond our field of view, the CND's hot southern lobe \citep{Yusef-Zadeh2001,Mills2017}, and the Western Arc begins precisely along the projected path of the wind, suggesting that ionized gas along this arc may trace the interface between the wind and the CND.
As the ionized gas is carried by the rotation of the CND, it forms the extended Western Arc.
This part of the minispiral is an ionized inner edge of the CND  that extends for approximately $\gtrsim50\%$ of a rotation \citep{Zhao2009, Zhao2010,Zhao2016}.

\subsection{Evidence of the Northern Lobe}

In the complementary north-northeast (NNE) direction, the CND contains considerably less material \citep{Hsieh2021}.
It has a more fragmented structure and an unfavorable orientation, making the observations of the NNE cone and the expected ionized features at its end less likely to be visible.
The Sgr~A~East supernova remnant \citep{Zhao2016, Maeda2002,Ehlorova2022,Zhang2023} complicates it even further by disrupting gas in this region and obfuscating evidence of the counter wind.
However, we expect the wind in the northeast to NNE direction to be present. 
There is a clear lack of cold molecular gas emission in this direction, a low-opacity cavity \citep{Schodel2010}, and an excess of X-ray emission \citep{Baganoff2003,Wang2013} (Figures \ref{fig:CO_Xray} and \ref{fig:Extinct}, and Appendix \ref{app:wind energy}).
The \citet{XRISM2025} indicate that the plasma in this direction is overionized (ions are stripped of more electrons than in collisional ionization equilibrium), consistent with activity from \sgra.
The velocity structure of the CND is such that the most blueshifted part of the CND aligns with the SSW cone, while the most redshifted gas aligns with the NNE direction (Figure \ref{fig:CND_vels}).

\subsection{Ionizing Radiation in the Central Parsec}\label{app:ionization}

The nuclear star cluster undoubtedly contributes a significant amount of ionizing photons to the region.
The particularly energetic young stellar cluster contributes $\sim 1\times10^{50}$~photons~s$^{-1}$  to the nuclear region, as estimated by \cite{Zhao2010}.
This would completely ionize 300 \Msun\ of molecular gas in $\sim60$ years, assuming idealized perfect absorption and ionization.
However, numerous recent works unambiguously demonstrated molecular gas presence inside the central parsec \citep[][and references therein]{Moultaka2015, Ciurlo2016,  Mills2017,  Zadeh2017a,  Moser2017, Goicoechea2018_HVMC, Goicoechea2018_HIFI, Hsieh2019, Hsieh2021}. 
The central pc has modest local dust opacities, along with a solar to supersolar metallicity \citep{Ciurlo2016,Schodel2010,Fritz2011,Vermot2025}, which would lead to much longer survival times of molecular gas.
Photoevaporation times are estimated to be $10^4-10^5$ years in such environments \citep[][and references therein]{Moultaka2009,Nakatani2019,Goicoechea2018_HVMC}.
It should be unsurprising that molecular gas survives in this environment.

Ionizing the Western Arc would be a challenge in this environment.
The local extinction in the central pc varies, but is about $A_{Ks}\simeq1$ magnitude \citep{Ciurlo2016,Schodel2010,Fritz2011}.
This is equivalent to an $A_v$ of 15, estimated using a power-law extinction curve with an index of 2.07 \citep{Fritz2011,Ciurlo2016}.
Given $A_v=1.086\tau_v$ and the radiative transfer equation $I=I_0e^{-\tau}$, where $\tau$ [$\tau_v$] is the optical depth [in the optical band], we find that only $\sim 1\times10^{43}$ photons s$^{-1}$ are free to escape the central pc.
The Western Arc makes up $\lesssim10$\% of the ionized gas luminosity of the minispiral \citep{Zhao2010} and the inner edge of the CND.
This means that the young stellar cluster produces an insufficient amount of ionizing photons to ionize the Western Arc of the minispiral.


However, if there were an additional directed source of radiation and mechanical energy, it would be capable of clearing a cavity, which would allow more ionizing radiation to reach, and ionize, the inner edge of the CND.
We identify just such a cavity in Figure \ref{fig:CO_Xray}.
We also stress that the direction of the black hole wind we argue for here is coincident with low-opacity regions in the the southern cavity of the CND identified in the infrared by \citet{Muzic2007}, \citet{Schodel2010} (Figure \ref{fig:Extinct}), and \citet{Lau2013}.
The inner edge of the minispiral is not equally ionized in all directions. 
The Western Arc of the minispiral begins where the outflow cone meets the CND and slowly decreases in intensity, temperature, and electron density as it rotates around the CND \citep{Zhao2010,Tsuboi2017}.
This is consistent with a low-opacity cavity where ionizing radiation can freely reach the inner edge of the CND.

In our interpretation, the \sgra\ wind clears a cavity through which ionizing radiation from the Nuclear Star Cluster and the black hole itself can more easily escape the inner parsec of the Galactic Center region, allowing more ionizing photons to reach the Western Arc of the Galactic Center minispiral. 

\subsection{Assessing a Supernova Origin}

We can also rule out a recent supernova origin of the cavity.
The last supernova in the Galactic Center occurred $10^4$ years ago and 5~pc away from \sgra \, \citep{Ehlorova2022,Maeda2002}, resulting in a short burst of activity from \sgra\ $\sim10^2-10^3$ years ago, and is too far to have caused the cavity.
Additionally, the possibility of a supernova in the CND has been investigated  \citep[e.g.,][]{Barna2025}.
In their simulation, a supernova would create a cavity in the CND lasting for $\sim10^3$ years, but would have provoked \sgra\ into a nonquiescent state for $\sim10^5$~yrs.
Whereas \sgra\ is known to be quiescent.

\subsection{Predicted Radio Luminosity and the Non‑Detection of Limb‑Brightening}
One would expect that the edge of the cavity created by a black hole's wind to be ionized. 
Unfortunately, detecting this would be quite challenging. 
The gas in the region is quite thin.
 The very fact that molecular gas through which the cavity carves its way was detected only now demonstrates how little of it there is.
While limb-brightening is always expected at the edges of such a cone, the amount of limb-brightening material is a small fraction of non-ionized material. 
The predicted flux density from the $\rm{P_{cavity}-P_{radio}}$ relationship is between $0.04-0.9$~Jy \citep{Cavagnolo2010, Birzan2008,Birzan2004} (assuming a spectral index of -0.8 to extrapolate the 1.4~GHz luminosity to 5.5~GHz). 
Such emission is readily accounted for by the variable $0.8\pm0.4$~Jy~\sgra\ point source \citep{Zhao2016}.
Additionally, there are multiple ionized gas structures in the region and ionized material along the line of sight \citep{Wang2010,Zhao2016}. 
Thus the emission from weak limb‑brightening, if present, would be (nearly) impossible to see in the existing observational data. 
Dedicated deep observations might be able to detect this.

\section{Properties of the Wind}

\subsection{Asymmetry}

While black hole outflows are expected to be symmetric close to the black hole, they can deflect due to the collisions with ambient gas in their path: the outflows follow the path of least resistance. 
In extreme cases, if one outflow runs into an obstacle but not the other, it can perish, whereas the other outflow can survive, resulting in an apparently one-sided outflow.
If an outflow is not powerful enough to drill through an obstacle head-on, it goes around it, with the outflow shape affected by the distribution of the surrounding gas. 

If the black hole wind/jet/outflow is relatively weak, the surrounding gas can deflect it producing an asymmetric, bent, or even one-sided structure.
Such behavior has been seen in both simulations \citep[e.g.][]{Kwan2023,Liska2023,Borodina2025,Lalakos2025} and observations \citep[][and references therein]{Veilleux2005, Garcia-Burillo2019, Bogensberger2024}.


The misalignment between the clearly identifiable SSW cone and the tentative counterwind cone plotted with dashed lines in Figure \ref{fig:Extinct}, can be an example of just such a case.
\citet{Eckart1992,Melia1996,Muzic2007} suggest that the minicavity is created from a collision with a fast (1000 \kms, $10^{38}\,\erg \, \sec^{-1}$) wind from the inner 0.2~pc.
Conversely, the bow shock structure of IRS 7 \citep{Yusef-Zadeh1992}, and the northern non-thermal filament \citep{Morris2017}, may also be an indicator of the existence of a northern counterpart of the wind, at least at some point in the past.

\subsection{Activity}

Simulations of the Galactic Center region showed that the CND actively ``rains''  molecular gas onto \sgra \ even in the presence of stellar winds, e.g., \cite{Solanki2023, Blank2016}.
\citet{Solanki2023}'s simulations show that the inner edge of the CND is collapsing inward, and the amount of molecular gas inside the CND is constantly growing and can fill the center completely.
There is  also ample observational evidence \citep[e.g.,][]{Goicoechea2018_HVMC, Hsieh2017, Tsuboi2018, Tanaka2021}, though the infall rates are order-of-magnitude estimates, inflow rates are consistently $\sim 10^{-3}~\Msun/\yr$.
We expect that the infall of molecular gas from CND can efficiently erase signatures of past activities on a timescale comparable to CND's orbital timescale. 
Numerical simulations are required for a more accurate estimate. 

We estimate that there is $\sim 300$~\Msun\ of the molecular gas within a 0.5~pc radius around \sgra (Appendix \ref{app:wind energy}), which is about the same as the amount of the cool ionized $10^4$ K gas in the minispiral \citep{Lo1983}.

So for the edges of the conical clearing to be as pronounced as they are, the wind must be presently active. 
Note that the clearing does not have to be devoid of gas completely.
It needs to be devoid of $\sim 100$ K molecular gas to appear empty in the image. 
We argue that the gas is partially evacuated and partially heated (Section \ref{app:ionization} \& Appendix \ref{app:wind energy}).



\subsection{Energetics of the Wind}\label{sec:analysis}

We estimate the energetics of the \sgra \, wind (or ``jet power'') by evaluating the power the wind needs to contribute to the overall energy balance of the inner Galactic Center region to keep the wind cone clear of cold molecular gas and to ionize the section of the minispiral's Western Arc (see Appendix \ref{app:ionization} for details of the calculations). 
We find the wind/jet power into the cleared cone to be $\sim 10^{37} - 10^{38}\,\erg \, \sec^{-1}$. The ratio of the Bolometric luminosity of \sgra \, ($\sim 3 \times 10^{36}  \, \erg \, \sec^{-1}$) to its jet power is similar to the one of M87$^*$ \citep{Genzel2010,EHT2021,Prieto2016}.

The wind/jet power estimated is larger than what can be provided by the stellar winds in the Galactic Center \citep{Calderon_2020ApJ...888L...2C,2020MNRAS.492.3272R,paumard_massive_stars}. Additionally, the winds tent to output the energy in more spherical geometry. These two observations together are ruling out stellar winds as the origin for the conical clearing.



\subsection{Lifetime of the wind in the SSW direction}

We can estimate the lifetime of the wind to be at least $\sim 2 \times 10^4$ years, which is approximately a quarter rotation of the CND.
It would take $\sim 2 \times 10^4$ years for the ionized gas at the end of the \sgra's wind to be carried by Keplerian rotation of the CND to form the full extent of the Western Arc \citep{Zhao2009,Zhao2010}, which is in reasonable agreement for the lifetime of the molecular gas within the central pc. 
For more accurate estimations of the lifetime of the wind, observations beyond 1 pc of the central region of the black hole are required.
We note that the local changes in the amount of ionized gas along the Western Arc could record past elevated activity of \sgra, or a combination of this with the local density variations.

\subsection{Orientation in the wind}

Orientation of the wind in the SSW direction is easy to identify more precisely. Assuming the center of the wind cone lies in the plane of the CND, and given the known inclination of the CND of $\sim$60 degrees relative to the plane of the sky \citep[e.g.,][]{Christopher2005,Requena-Torres2012,Martin2012,Lau2013,Ukani2025}, the de-projected center of the outflow has an inclination of $\sim70$ degrees to the line of sight and points away from the observer.
Because the wind cone is relatively wide, part of the wind is still pointing toward the observer along the line of sight.

As the distribution of gas in the Galactic Centers changes on the galactic timescale, the wind could change direction as well. 
So the wind/jet energy will be deposited more broadly around the central region of the Galaxy than if a wind/jet keeps one preferred direction.

\section{Conclusion}

In this work, we combined $\sim$5 years of ALMA observations to produce the deepest and highest‑resolution map, 100 times deeper and 80 times sharper than previous CO maps, of the cold molecular gas within $\sim$1 pc from \sgra\ to date.
The image reveals the presence of cold molecular gas around the black hole and inside the inner radius of the surrounding molecular ring (CND).

On the molecular gas map we identify a large conical clearing, which we argue is the path of a hot active wind from \sgra\ (Figure \ref{fig:CO_Xray}).
The wind is not perpendicular to the Galactic plane, nor is it perpendicular to the CND, which is the dominant supplier of the material in the region. 
The wind is relatively weak and likely deflected from its launch orientation by the interaction with the ambient gas, as evident from the asymmetry of the cavities it creates.
It is likely that the wind/jet changes direction and thus deposits energy into the Galactic Center region more broadly than a wind/jet that keeps one preferred direction.
The findings in this work are consistent with intricate and tangled phenomena of feeding and feedback in  the Galactic Center revealed in previous studies.
Our finding also is a significant step towards resolving the long-standing mystery of \sgra's missing wind and delivers the most detailed picture yet of black hole feeding and feedback in the center of the Milky Way.

In the broader picture, \sgra\ is a typical example of underfed quiescent SMBHs, which is the dominant phase, by time spent in it, of SMBH evolution in the Universe. 
We therefore would argue that the phenomena observed here of weak and wandering feedback over the central region of the galaxy are not unique and likely apply to most of the other quiescent galaxies in the Universe.



\begin{acknowledgments}

We are grateful to Claire J. Chandler for advice on data processing, and to Ivan Marti-Vidal, Anna Ciurlo, Mark Morris, Shoko Sakai, Tuan Do, Roger Blandford, Nick Scoville, Sasha Tchekhovskoy, Rick Perley, and Elizabeth Mills for comments and discussions, to Pei-Ying Hsieh for sharing CS maps, and to Jurgen Ott, Jin Koda, Eliot Quataert, and Sean Ressler for contribution to writing ALMA proposals, and to the anonymous referee for their helpful comments.

M.D.G. is supported by the CIERA Postdoctoral Fellowship from the Center for Interdisciplinary Exploration and Research in Astrophysics at Northwestern University.

This paper makes use of the following ALMA data: ADS/JAO.ALMA \#2012.1.00543.S,
\#2016.1.00870.S, \#2017.1.00995.S, \#2017.1.00040.S, and \#2019.1.01559.S.
ALMA is a partnership of ESO (representing its member states), NSF (USA) and NINS (Japan), together with NRC (Canada) and NSC and ASIAA (Taiwan) and KASI (Republic of Korea), in cooperation with the Republic of Chile. 
The Joint ALMA Observatory is  operated by ESO, AUI/NRAO and NAOJ.

The National Radio Astronomy Observatory is a facility of the National Science Foundation operated under cooperative agreement by Associated Universities, Inc.

This research has made use of data obtained from the Chandra Data Archive provided by the Chandra X-ray Center (CXC).

This work used computing resources provided by Northwestern University and the Center for Interdisciplinary Exploration and Research in Astrophysics (CIERA). This research was supported in part through the computational resources and staff contributions provided for the Quest high performance computing facility at Northwestern University  which is jointly supported by the Office of the Provost, the Office for Research, and Northwestern University Information Technology.
  
\end{acknowledgments}





%
\facilities{ALMA}

\software{astropy \citep{Astropy2013,Astropy2018,Astropy2022},  
          UVmultifit \citep{marti-vidal2014}, 
          CASA \citep{CASATeam2022PASP}
          }

\appendix

\begin{figure*}[!htb]
\centering
\includegraphics[width=0.99\textwidth]{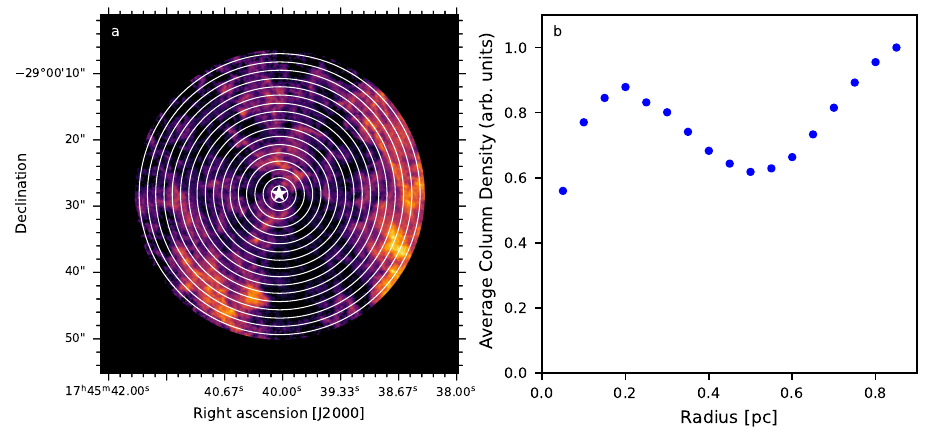}
\includegraphics[width=0.99\textwidth]{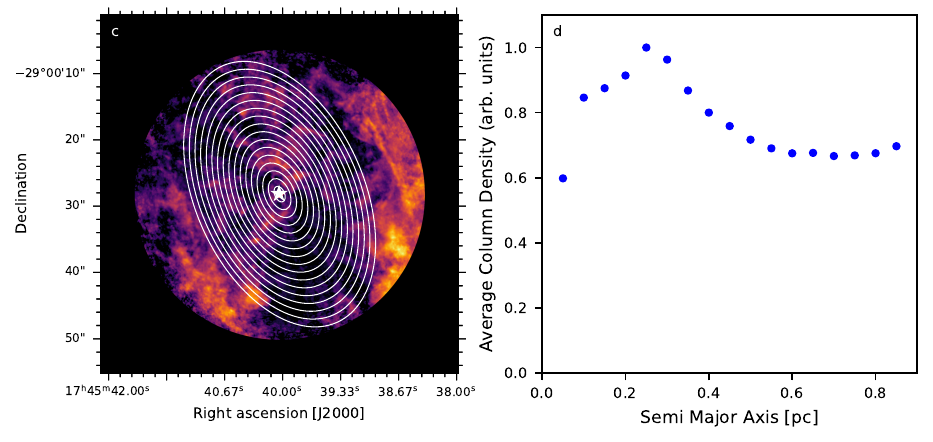}
\caption{
Surface density profile of the \bco \, emission.
\textbf{a,} Consecutive concentric circles used to measure the average column density profile surrounding \sgra (white star). 
\textbf{b,} Average column density of gas extracted within each circle in arbitrary units as a function of radius from \sgra.
\textbf{c,} Consecutive concentric ellipses with an axial ratio of 0.6 and a position angle of 20 degrees east of north, corresponding to inclination and orientation of the CND on the sky \cite{Martin2012}. 
\textbf{d,} Average column density of gas extracted within each ellipse in arbitrary units as a function of radius from \sgra.
}
\label{fig:annuli_density_prof}
\end{figure*}

\begin{figure*}
    \centering
    \includegraphics[width=0.49\textwidth]{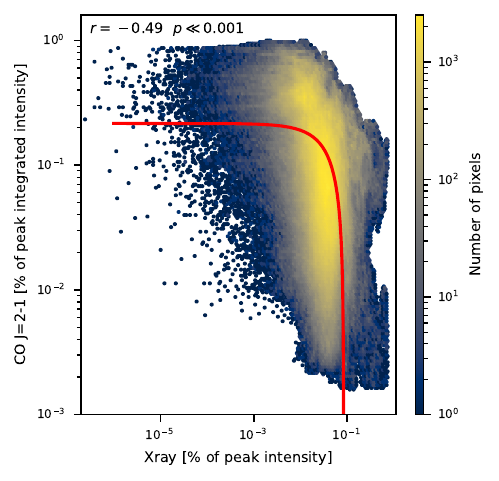}
    \includegraphics[width=0.48\textwidth]{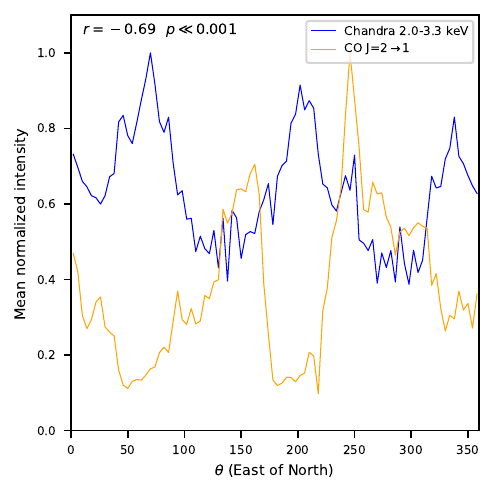}
    \caption{(Left)  Hexagonally-binned spatially-corresponding \bco\ integrated intensity and 2.0-3.3 keV X-ray intensity pixels inside the ALMA primary beam (Figure \ref{fig:CO_Xray}).
    The red line shows a linear fit to the data weighted by the number of pixels in each hexbin. 
    While there is a wider range of X-ray values than CO values, the fit reveals an anticorrelation. 
    The Spearman correlation test also indicates a modest anticorrelation and rejects the null hypothesis.
    (Right) Compares the azimuthal profile over the ALMA primary beam of the \bco\ and 2.0-3.3 keV X-ray maps.
    While there is a broad low level of x-ray and CO emission across the map, the strongest emission is inversely correlated. 
    Spearman correlation coefficients indicate a strong anticorrelation and reject the null hypothesis. }
    \label{fig:CO_Xray_Corr}
\end{figure*}

\begin{figure*}[!htb]
\centering
\includegraphics[width=0.75\textwidth]{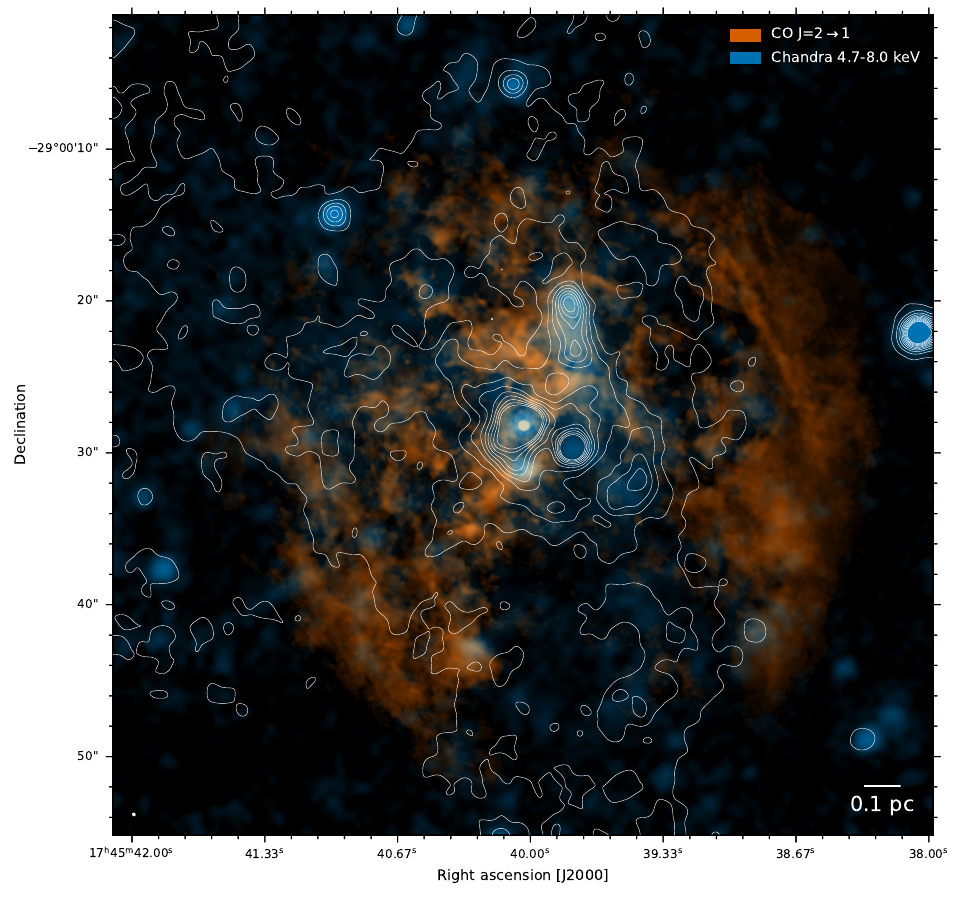}
\caption{
Molecular gas and 4.7-8.0 keV X-ray emission around \sgra.
ALMA's \bco\ integrated-flux map (orange). Chandra's 4.7-8.0 keV X-rays (blue) primarily trace point sources \citep{Morris2010}. 
White contours indicate 2.0-3.3 keV X-ray emission from Figure \ref{fig:CO_Xray} with 1.26$\times10^{-8}$ photon~s$^{-1}$~cm$^{-2}$ steps until 1.26$\times10^{-7}$ photon~s$^{-1}$~cm$^{-2}$.
}
\label{fig:CO_Xrayhigh}
\end{figure*}

\section{Comparison with known structures}\label{app:known_structures}
\begin{figure}[!htb]
    \centering
    \includegraphics[width=0.99\textwidth]{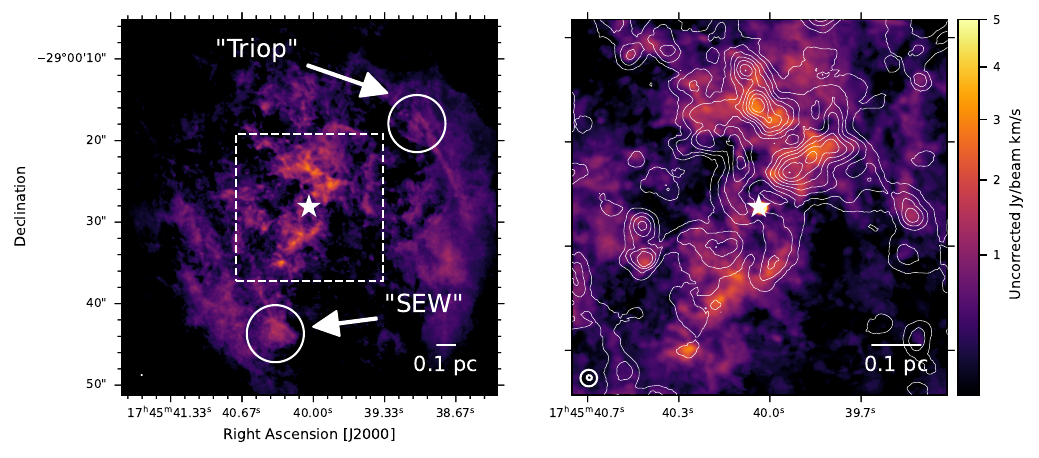}
    \caption{ 
    (Left) Known molecular gas structures within the CND visible on \bco \, map.
    \textbf{Primary-beam-uncorrected} \bco\ integrated-flux map showing the ``SEW'' and ``triop'' structures in the CND \cite{Moser2017}. 
    The structures are indicated with arrows.
    The white star marks the location of \sgra.
    The square indicates the field of view of the right pannel.
    (Right) Known molecular gas streamers within 0.5~pc of \sgra  revealed in \gcs\ emission with an angular resolution of 0.7\arcsec$\times$0.4\arcsec \citep{Hsieh2019,Hsieh2021}.
    Contours show the  0.15, 0.46, 0.78, 1.10, 1.41, 1.71, and 2.03 Jy beam$^{-1}$ levels of the highly excited \gcs\ line.
    There is good agreement between the gas streamers.
    The Spearman correlation test reveals a weak correlation.
    However, this is expected as \gcs\ is excited at higher temperatures and densities, so low-density, cooler parts of streamers will not be revealed in \gcs.}
    \label{EDfig:MoserStructures}   
\end{figure}

\begin{figure}[!htb]
\centering
\includegraphics[width=0.92\textwidth]{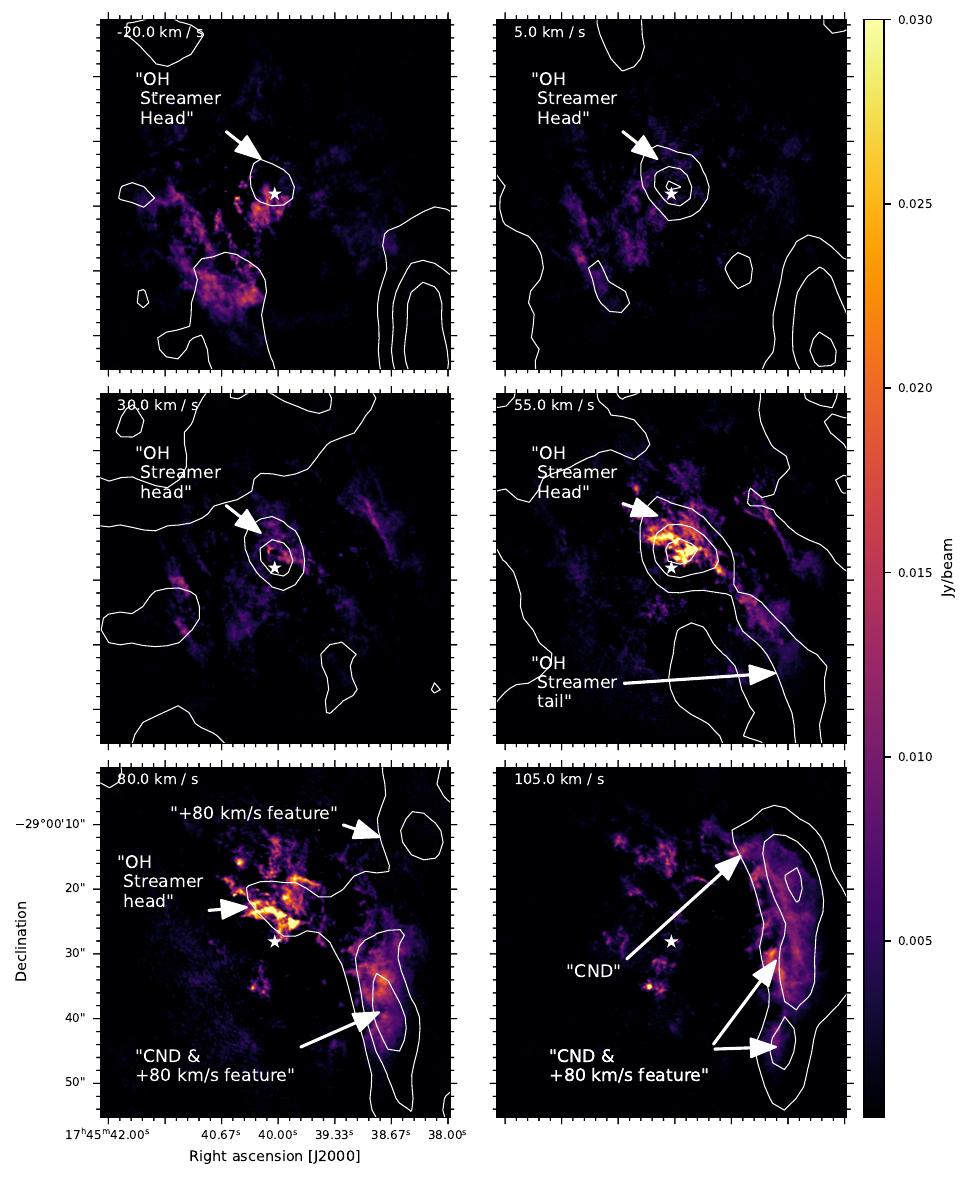}
\caption{
    Known OH streamer visible on \bco \, map.
    Channel maps of the \bco\ line (presented in work) and the OH 1.667~GHz line belonging to the OH streamer  \cite{Karlsson2015}. 
    The OH line is shown with white contours at 50, 100, and 150~mJy~beam$^{-1}$. 
    The head and tail of the molecular streamer are labeled, and the 80 \kms\ feature from \citet{Karlsson2015}.
    The OH line shows the edge of the CND in the 105~\kms\ channel. 
    The position of \sgra\ is marked with a white star.}
\label{fig:CO_OH_Channels}
\end{figure}

\begin{figure}[!htb]
\centering
\includegraphics[width=0.96\textwidth]{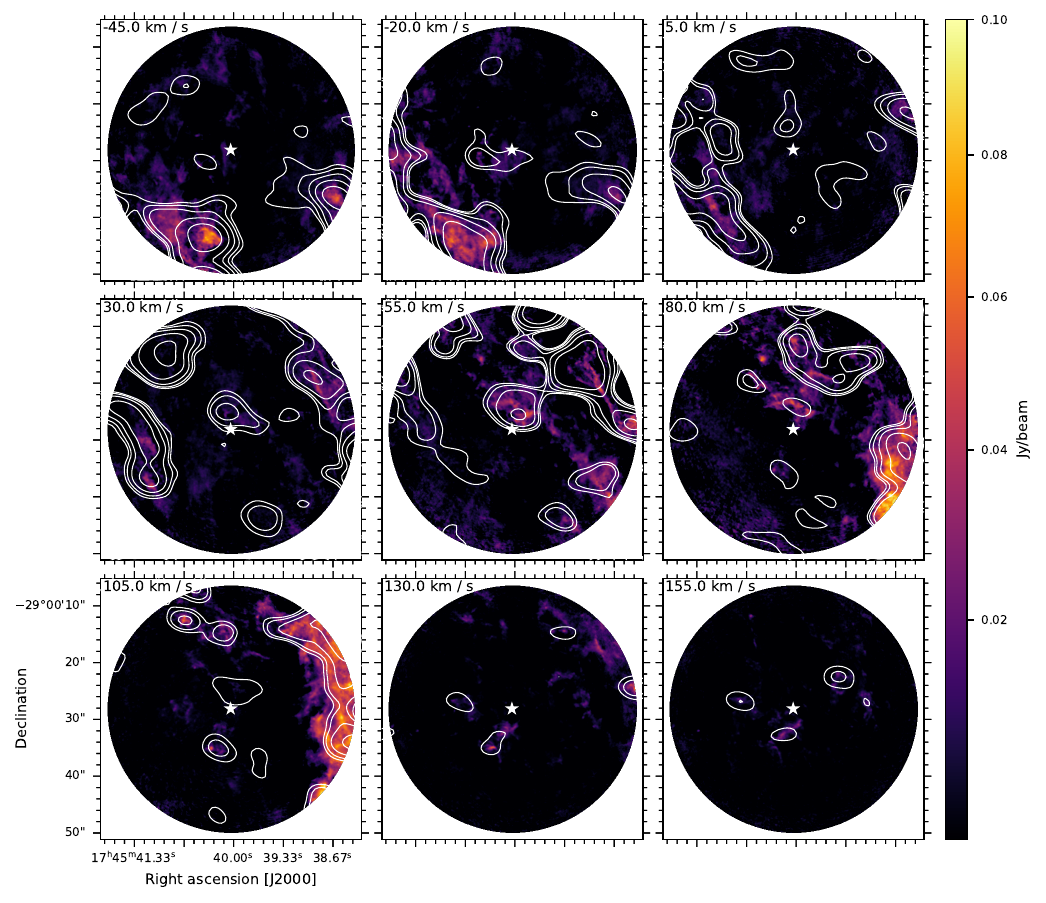}
\caption{
    Comparison of the primary beam corrected channel maps of the \bco\ line (presented in work, color) and the \dhcn\ (contour) ALMA 7m maps \citep[ 4.8\arcsec$\times$3.0\arcsec\ resolution, position angle 80\deg, ][]{Martin2012}. 
    The \dhcn\ line is shown with white contours at 5, 10, 20, and 40 times 0.1 Jy. 
    \dhcn\ is also an established tracer of the CND  \citep{Martin2012,Montero-Castano2009,Guesten1987} but less susceptible to Galactic foreground, as the +80 \kms\ cloud is less apparent. 
    \bco\ and \dhcn\ lines are tracing gas of somewhat different conditions, so they cannot exactly match each other. 
    Nonetheless,the channel maps are in excellent agreement, demonstrating that the \bco\ line is indeed tracing the CND and central parsec.
}
\label{fig:CO_HCN43_Channels}
\end{figure}

A significant concern in observing the Galactic Center is obscuration by the foreground disk of the Milky Way. 
That is, gas residing between our solar system and \sgra\ may block our view of  \sgra\ and the surrounding Galactic Center structures. 
\citep{Binney1991} identified absorption features at vLSR from -121~\kms\ to {-145}~\kms\ attributed by foreground gas.
This absorption originates from non-circular orbits of the Galaxy’s barred potential.
Indeed, our \bco\ channel maps show a lack of emission at these velocities.
Beyond this range, however, there appears to be no major evidence for Galactic foreground across the ALMA primary beam.
\cite{Gorski2025} show that foreground absorption towards the CND in this field of view at these frequencies is negligible.
Moreover, we detect known structures in the CND:
(i) The CND edge and the velocities within the CND, as measured by HCN \citep{Martin2012, Hsieh_2018}, are well-matched and reproduced in our data (Figure \ref{fig:CND_vels}B). 
(ii) The ``triop''\citep{Moser2017}, a triangular crayfish-shaped structure northwest of \sgra is clearly visible (Figure \ref{EDfig:MoserStructures}).
(iii) The southern extension west (SEW)\citep{Moser2017} is observed just south and slightly east of \sgra\ (Figure \ref{EDfig:MoserStructures}).
(iv) Known streams of molecular gas within the central 0.5 pc \citep{Hsieh2019} are clearly visible.
Figure \ref{EDfig:MoserStructures} shows nuclear filaments traced in by the highly excited \gcs\ line.
(v)  The OH streamer \citep{Sandqvist1989,Karlsson2003,Karlsson2015} is visible in our data.
Figure \ref{fig:CO_OH_Channels} shows correlations between the CO and OH gas.
The OH data cube has an angular resolution of 7\arcsec\ by 5\arcsec\ with a position angle of 31~degrees. 
\citep{Karlsson2015} shows that the OH streamer is a structure that exists within the CND and interacts with \sgra. 
Our CO data trace both the head and tail of the streamer, matching its structure and velocity.
We also compare with known tracer of the CND  HCN in emission \citep{Guesten1987, Montero-Castano2009, Martin2012}. 
\ahcn\ channel maps from \cite{Hsieh2021} and \dhcn\ maps \cite{Martin2012}, binned to 25 \kms \, spectral resolution, show exceptional agreement with the \bco\ maps presented here (Figures \ref{fig:CO_HCN_Channels} \& \ref{fig:CO_HCN43_Channels}).
\dhcn\ is less susceptible to intervening clouds such as the +80 \kms\ cloud. 

The agreement in CND velocities and the detection of nuclear structures such as the ``triop'', southern extension west (SEW), and OH streamer, confirm that we are genuinely detecting \bco\ gas within a radius of $\sim$1~pc surrounding \sgra.
Our observations are mostly unobscured by Galactic foreground, but gas infalling from the CND may yet obscure some of the nuclear structures.

Figures \ref{fig:CO_Xray} and \ref{fig:CO_Xray_Corr} show anti-correlation between \bco \, and $2.0-3.3$ keV X-ray emission of hot gas in the Galactic Center discussed in the main text. 
We can also compare the \bco\ integrated-flux map with the $4.7-8.0$~keV X-ray emission observed by Chandra \citep{Morris2010}.
The $4.7-8.0$~keV X-ray emission correlates more with point sources rather than with diffuse gas \citep{Morris2010,Zhu2018}.
We do not observe as strong of an anti-correlation of \bco \, emission with the point source distribution as with the diffuse gas distribution (Figure \ref{fig:CO_Xrayhigh}).

\section{Molecular gas mass in the inner 1 pc of the Galactic Center}\label{app:molecular}

To estimate the amount of molecular gas, we use the same method is in \citet{Gorski2025}.
Consider the part of the \bco\ map that overlaps with the CND between $0.5$~pc and $0.85$~pc from \sgra \, ($D =8$~kpc; Figure \ref{fig:CND_vels}b).
One can calculate that this region contains about 10\% of the CND both by area and by total molecular line emission (here we used HCN emission map \citep{Hsieh2021}, which is one of the most detailed maps of the CND). 
The total CND mass is $\sim 3\times 10^4$~\Msun \,\citep{Genzel2010,2015Harada_CND,James2021}, thus the mass within the overlapping region is $\sim 3 \times 10^3$~\Msun. 
At the same time, the ratio of \bco\ emission between 0.5 and 0.85~pc and inside the 0.5~pc is 25:7 (Figure \ref{fig:annuli_density_prof}a).
One may naively conclude that the amount of gas inside 0.5~pc should be $\sim 840$~\Msun. 
However, it is important to include the CND's optical depth. 
Using the radiative transfer code RADEX \citep{vanderTak2007} and average parameters of the CND, we find that average optical depth of the CND is $\tau \simeq 1,$ implying that only $e^{-\tau} = 0.37$ of the \bco \, emission from the dense parts of the CND reaches the observer.
At the same time, the inner 0.5 pc is filled with thin gas, with a likely optical depth of $\ll 1$.
So instead of the $25:7$ ratio, we should use $68:7$. 
Consequently, the amount of the molecular gas inside a 0.5 pc radius around \sgra \ should be $\sim 310$~\Msun.
Assuming a cylindrical volume with a radius of 0.5~pc and a height equal to the CND’s vertical extent (1~pc), we find an average gas density of $\sim 10^4 \, \cm^{-3}$.
It is clear from our image (Figure \ref{fig:COpb_vel}) that the molecular gas is not uniformly distributed in the region, but largely forms clumps and streamers that are consistent with the nuclear streamers indicated by \citet{Hsieh2019} in their position-angle vs. velocity analysis, so the actual density fluctuates a few orders-of-magnitude about this value.
This method for calculating gas mass relies on the integrated knowledge of the CND accumulated over decades of observations, and therefore, in our opinion, the most reliable way to estimate the gas mass in the region.

This estimate is just lower than the lower-density regions of the CND, where densities range between $2\times10^4 \, \cm^{-3}$ and $1\times10^6 \, \cm^{-3}$ \citep{James2021,Requena-Torres2012}. 
This aligns with simulations of the CND \citep{Solanki2023} that show that  the CND ``rains'' molecular gas into the Galactic Center through the infall of clouds and streamers.

A cold molecular gas phase is likely dominant within the inner $\sim 0.5$ pc around \sgra, as the density of the hot X-ray-emitting gas is only $\sim (1-2) \times 10^{2} \, \cm^{-3}$ \citep{Quataert2002}.
The molecular gas phase is expected to dissipate near \sgra, but unfortunately, a large redshifted gas streamer currently contaminates our view of the innermost radii.

\begin{figure}[!htb]
\centering
\includegraphics[width=0.95\textwidth]{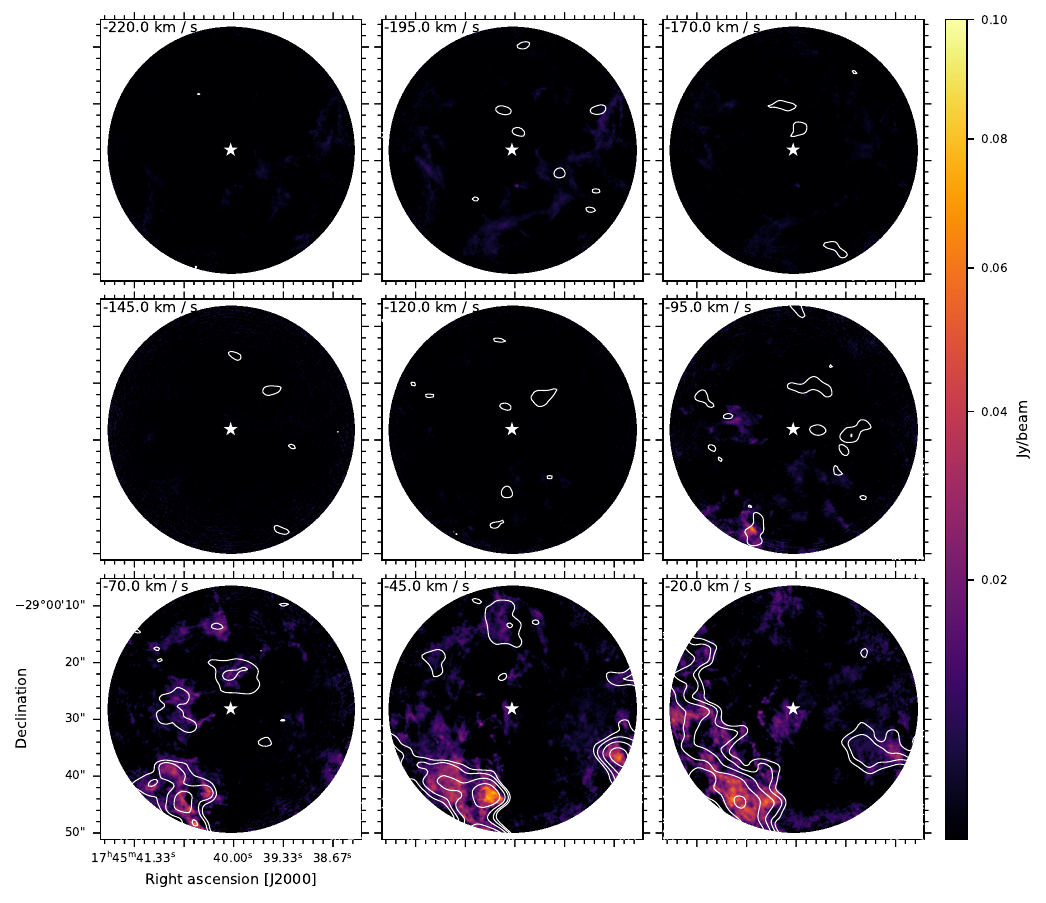}
\caption{
    Comparison of the primary beam corrected channel maps of the \bco\ line (presented in work, color) and the \ahcn\ line (contour). 
    The HCN line is shown with white contours at 3, 6, 12, 24, and 48 times 0.01 Jy. 
    HCN is an established tracer of the CND \citep{Christopher2005,Martin2012,Montero-Castano2009,Guesten1987}.
    \bco\ and \ahcn\ lines are tracing gas of somewhat different conditions, so they cannot exactly match each other. 
    Nonetheless, the channel maps are in excellent agreement, demonstrating that the \bco\ line is indeed tracing the CND and central parsec. 
}
\label{fig:CO_HCN_Channels}
\end{figure}

\begin{figure}[htbp]
    \ContinuedFloat 
    \centering
    \includegraphics[width=0.96\textwidth]{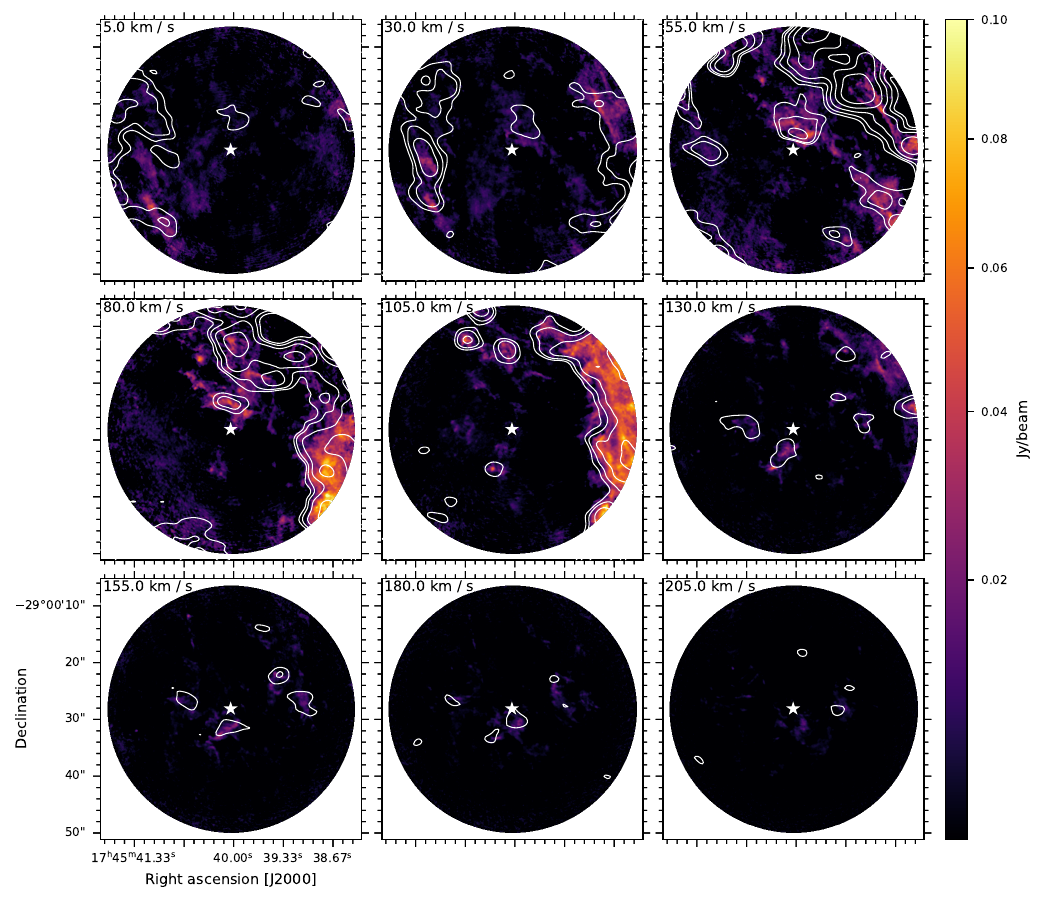}
    \caption{Figure \ref{fig:CO_HCN_Channels}  continued)}
\end{figure}

\section{Estimate of the Wind Energetics}\label{app:wind energy}

In this section we estimate the energy budget of the wind, or ``jet power'', by considering multiple effects. \textbf{We first calculate the total energy budget required to heat and move the gas in the region, in order to then evaluate from which sources this energy can be supplied.}
Let us start by estimating the amount of energy required to heat a portion of the CND and form the Western Arc of the minispiral. 
The total mass of the minispiral is about 340~\Msun, and it consists of three parts: Northern Arm, Eastern Arm, and Western Arc \citep{Mini_Spiral_Kunneriath,Lo1983,Zhao2010}. 
Assuming that the full Western Arc has a mass of about 100~\Msun, we observe
that only about one-third of the Western Arc lies within the cone of the wind. 
To heat $\sim 33$~\Msun\ of gas from a few 100~K (typical in the CND) to $\sim 10^4~\K$ (typical in the minispiral) over the duration the arm segment is passing through the conical wind requires about $\sim 4.4 \times 10^{35}~\erg \, \sec^{-1}$. This is calculated from purely thermodynamic consideration and represents the energy rate that needs to be supplied to the given amount of gas but not accounting for the losses (e.g., cooling), which are discussed later. The gas is treated as predominantly hydrogen.
We assumed the distance to the arm to be $1.5$~pc and the rotational velocity to be Keplerian velocity at this radius. 
In comparison, stellar winds produced by the disk of Wolf-Rayet stars located at $\sim 0.15$~pc from the black hole \citep{10.1093/mnras/sty1146,Calderon_2020ApJ...888L...2C}
are able to deposit up to $\sim 1 \times 10^{35} ~\erg \, \sec^{-1}$ into the \sgra's wind cone \citep{Solanki2023}.
Thus, stellar winds alone are insufficient to produce this effect.


The extent of the arm and survival time of the molecular gas indicate that the current wind was active for at least $\sim 2 \times 10^4$ years.
The dominant component of the wind energetics, however, is likely used to keep the wind cone clean, either by clearing the molecular gas (i.e., moving it out of the cone) or by heating it to $\gtrsim 1,000$~K, at which emission of \bco\ rotational transition is suppressed.

Previously, we estimated that there is about $3 \times 10^3$\,\Msun\ of cold CND gas between the $0.5$ and $0.85$~pc radii from the black hole.
Assuming that this is an average amount of gas between these radii, we find that about 250~\Msun\ of this gas would be passing through the wind cone every $\sim 2200$~years. We used the projected size of the observed cone, i.e., 45 degrees, and Keplerian velocity of rotation, which is a good fit for the velocities of the gas in the CND (e.g., \citealt{Ukani2025}). To calculate Keplerian velocities, we use the joint potential of the \sgra \, and the nuclear star cluster (Galactic Center potential) explicitly given in \citealt{Chatzopoulos_2015}, which also provides the mass distribution in the Galactic Center with radius.
The amount of energy required to move this amount of material in the Galactic Center potential from $\sim 0.7$~pc to $\sim 1$~pc from \sgra\ is $\sim 4 \times 10^{38}~\erg \, \sec^{-1}$. We calculate this using the difference in the potential energies. 
On the other hand, to heat the same amount of molecular gas to $10^3$ K (or $10^4$ K \citep{Maloney1996}), at which the emission of \bco\ would largely be suppressed (or CO would fully dissociate), requires $\sim 8 \times 10^{35}~\erg\, \sec^{-1}$ (or $\sim 10^{37}~\erg\, \sec^{-1}$).

For the gas within the inner 0.5~pc, we estimate that less than $\sim 25$~\Msun \, lies within the wind cone.
To evacuate this gas from about $\sim0.2$~pc to about $\sim 1$~pc radius from the black hole in the Galactic Center potential requires about $4 \times 10^{38}\,\erg \, \sec^{-1}$ (this is purely a coincidence that this matches the value above).
In contrast, heating this gas to $10^3$~K, to suppress emission of the \bco\ line would require just under $\sim 4\times 10^{35}\,\erg \sec^{-1}$.
Note that while the amount of gas inside 0.5~pc is less than the amount of gas between 0.5 and 0.85~pc, the gas spends less time inside the geometrical area of the wind cone  due to faster Keplerian rotation and the cone's smaller extent closer to the black hole.

In reality, evacuation and heating would contribute jointly.
Some fraction of the gas will be evacuated, and another fraction will be heated.
So we estimate the jet power to be about $10^{37} - 10^{38} \, \erg \, \sec^{-1}$. 
It is comparable to the ionizing energy and mechanical energy of stellar winds of the young cluster and WR stars in the Galactic Center. 
The WR stars and the young cluster would not be able to clear the wind cone(s) by themselves, as they distribute energy relatively spherically, but they would contribute to the total energy budget within the cavities. 
The winds and emission from IRS13 are also expected to contribute.

The calculations of the jet power (or the wind energy) conducted here are order-of-magnitude estimations. 
We do not attempt to calculate the energy required for the gas to escape from a few gravitational radii from the black hole. This calculation would be very sensitive to the exact origin of the wind and the mass of evacuated gas, which are unknown.
In general, accurate jet power calculations are only possible in simulations, as they can track the origin of the outflowing gas. 

Velocity of the wind is changing strongly depending on the distance from the black hole, as the wind is essentially bounded to the region and not escaping it. If one to attempt to calculate wind velocity, one can proceed in a few different ways. Assuming 100\% of the wind power goes into to the mechanical power of ejecting the wind, assuming the mass loss rate is equivalent to the accretion rate (this is close to the truth as only about few percents of the few$\times 10^{-6}~ \Msun~\rm{yr}^{-1}$ \citep[e.g.][]{Wang2013} gas captured at Bondi radius ($\sim 0.04$ pc) is reaching the black hole) we can estimate the velocity of the wind at Bondi to be $\sim1000-4000$ \kms. We can also estimate the velocity of the winds from the fact that the wind cleared cavity largely ends at around $\sim 1.5$ pc. The velocity of gas that can reach 1.5 pc away from the black hole and come to rest there should be $\sim 1000$ \kms \, at Bondi radius. Taking into account that this estimate does not include collisional or radiation losses, we conclude that the wind velocity should be $>1000$ \kms. Both estimations are consistent with each other and  consistent with predictions from \citep{Eckart1992,Melia1996,Muzic2007}.

Our estimation of Sgr A* jet power is about $10^2$ times larger than its bolometric luminosity $\sim 3 \times 10^{36}  \, \erg \, \sec^{-1}$ \citep{Genzel2010}, which is about the same ratio as for M87$^*$ \citep{EHT2021,Prieto2016}.

\bibliography{blackholes_and_sgra, moleculargas_and_galaxies, sgra, software}
\bibliographystyle{aasjournalv7}

\end{document}